\documentclass[]{spie}  

\usepackage{amsmath,amsfonts,amssymb}
\usepackage{graphicx}
\usepackage[numbers]{natbib}
\usepackage[colorlinks=true, allcolors=blue]{hyperref}

\newcommand{\um}{$\mu$m}

\title{Wide field Slitless Spectroscopy with JWST's MIRI}

\author[a]{Andreea Petric}
\author[b]{Sarah Kendrew}
\author[a]{David Law}
\author[a]{Alberto Noriega Crespo}
\author[a]{Bryan Holler}
\author[a]{Jane Morrison}
\author[a]{Jonathan Aguilar}
\author[a]{Misty Cracraft}
\author[a]{Ian Wong}
\author[c]{Stacey Alberts}
\author[a]{Dean Hines}
\author[c]{Kate Rowlands}
\author[a]{Ned Molter}
\author[a]{G. C. Sloan}
\author[a]{Mike Engesser}
\affil[a]{STScI, AURA}
\affil[b]{European Space Agency, Space Telescope Science Institute, Baltimore, MD, USA}
\affil[c]{AURA for ESA, Space Telescope Science Institute, Baltimore, MD, USA}

\authorinfo{Further author information: email apetric@stsci.edu}

\begin{document} 
\maketitle

\begin{abstract}
We present a snapshot of the ongoing efforts to obtain background-subtracted, wavelength-, and flux-calibrated spectra taken with the new Wide-Field Slitless Spectroscopy (WFSS) mode for the MIRI instrument on the James Webb Space Telescope (JWST), offered for the first time in JWST Cycle 5 (starting July 2026). We describe here the capabilities of the new mode, the operational concept, and an overview of the calibration and pipeline development activities that are currently ongoing. 




\end{abstract}

\section{Introduction}

\begin{figure}
    \centering
    \includegraphics[width=0.5\linewidth]{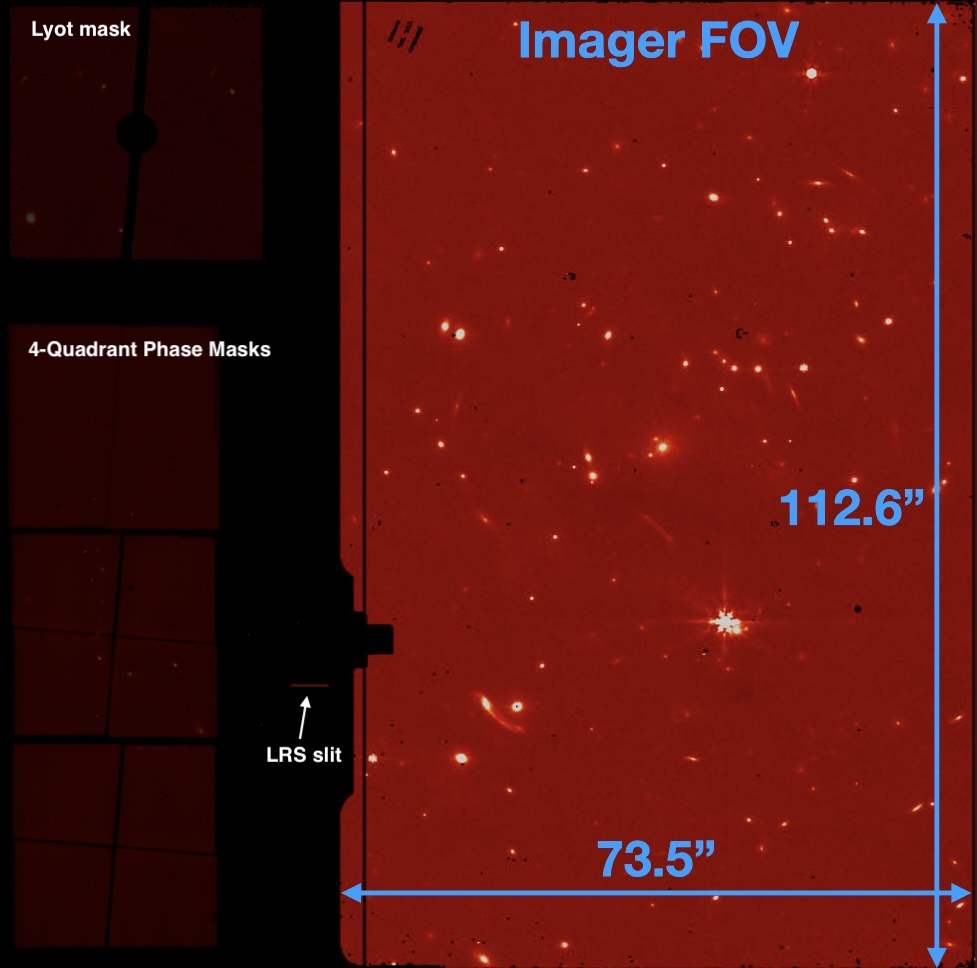}
    \caption{Overview of the Imager focal plane layout, showing the main Imager field, and key locations for the LRS and coronagraphic imaging modes. Figure adapted from \href{https://jwst-docs.stsci.edu/jwst-mid-infrared-instrument/miri-observing-modes/miri-imaging}{JWST Documentation}.}
    \label{fig:imager_fov}
\end{figure}

The MIRI instrument on JWST covers wavelengths from $\sim$5 to 28~\um\ with imaging, coronagraphic imaging, low-resolution slit and (single-object) slitless spectroscopy (LRS), and medium-resolution integral field spectroscopy (MRS)~\citep{2015PASP..127..665R, 2023PASP..135d8003W}. The instrument has 3 Si:As Impurity Band Conduction (IBC) detectors, 1032 $\times$ 1024 px; 2 are dedicated to the MRS mode, and 1 is used for imaging, coronagraphic imaging, and low-resolution spectroscopy~\citep{2015PASP..127..584R}. The imaging detector has a number of subarrays, dedicated regions for coronagraphy using three 4-quadrant phase masks and a Lyot mask, and a fixed slit for the LRS slit mode. The LRS mode~\citep{kendrew2015} uses a double prism mounted in the Imager filter wheel, which provides R$\sim$40-160 spectroscopy from $\sim$5 to 12~\um\ (the prism throughput extends to 14~\um, the transmission drops off steeply in the 10 -- 14~\um\ region). The Imager focal plane is shown in Fig. \ref{fig:imager_fov} with key locations marked. 

Starting in May 2024, a project has been ongoing to develop a Wide-Field Slitless Spectroscopy (WFSS) mode, using the LRS double prism in combination with the large Imager field of view to provide low-resolution dispersion of the full Imager field. This is the most significant new capability enabled for MIRI since the start of operations in 2022, and the only multiplexed spectroscopic capability in the mid-IR on JWST. It provides strong complementarity with similar WFSS modes on MIRI's near-IR counterparts on JWST, NIRCam and NIRISS. The mode was formally offered to the community for science observations from JWST Cycle 5 (July 2026).

In this paper we will describe the operations concept of the mode and some of the calibration and pipeline development work performed to enable science observations. 

\section{Operations Concept}

\begin{figure}
    \centering
    \includegraphics[width=0.55\linewidth]{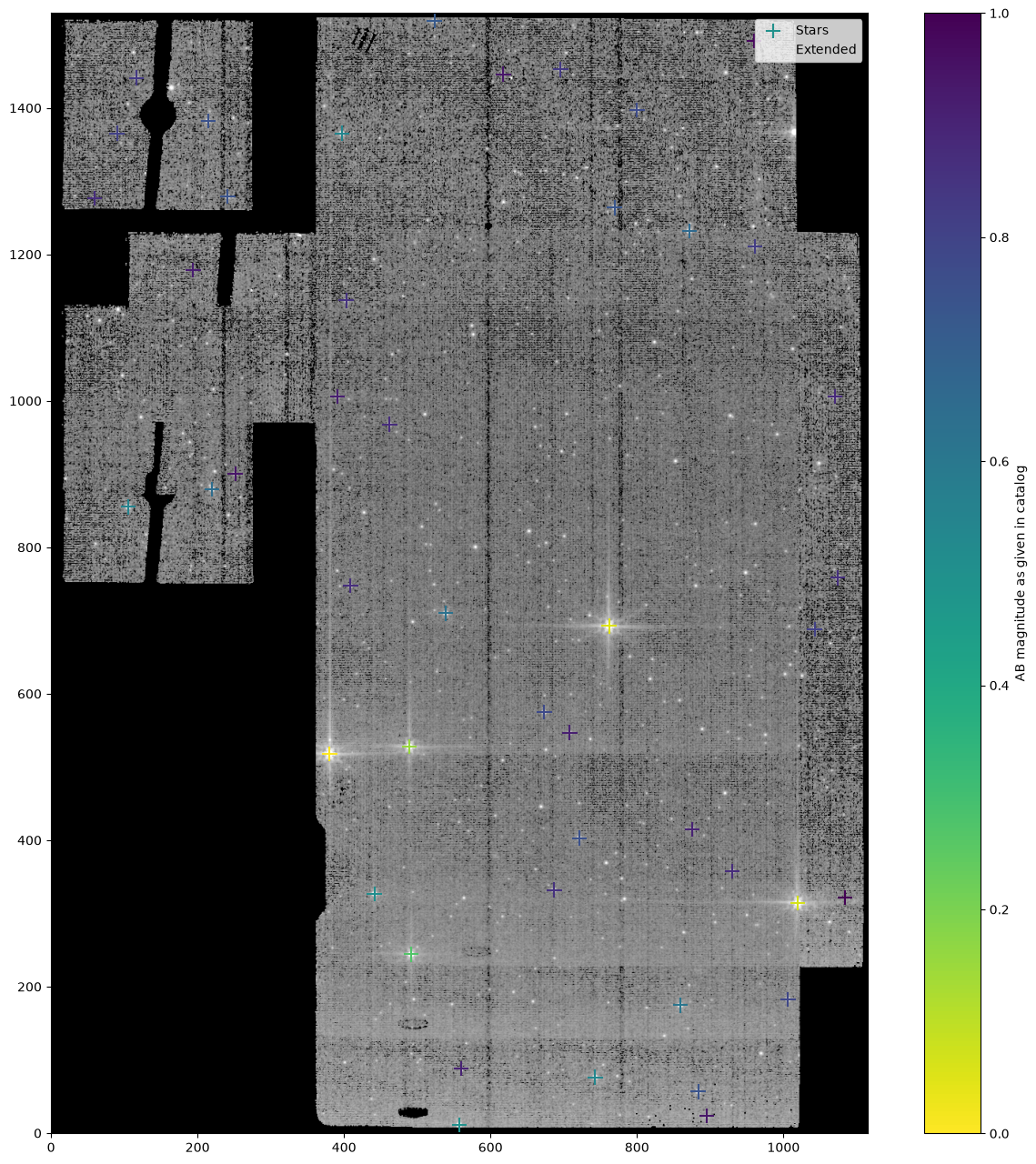}
    \includegraphics[width=0.44\linewidth]{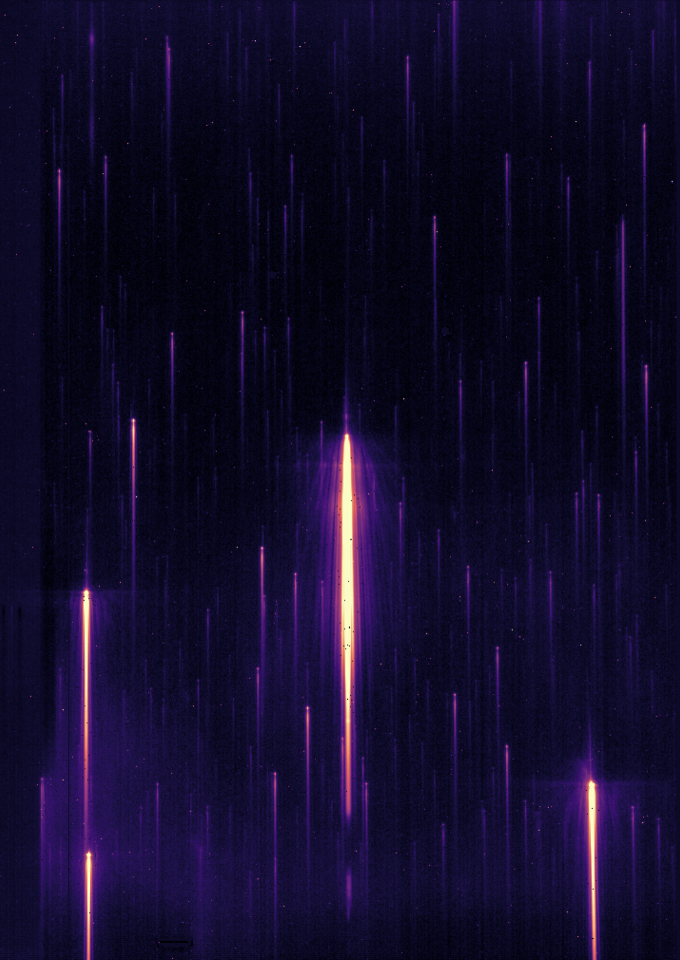}\\
    \caption{Examples of combined direct (left) and dispersed images from a MIRI Wide Field Slitless Spectroscopy Exposure taken as part of WFSS Science Verification program PID 9505.}
    \label{fig:dispSpec}
\end{figure}

Operationally, a WFSS observation starts with a direct image obtained with any of the F560W, F770W, F1000W, F1130W, F1280W, or F1500W filters, which will later be used to set the positions of the spectral extraction apertures. The imaging is then followed by a series of dispersed exposures with a minimum of 4 dither positions. The WFSS dithering uses the Cycling pattern developed for MIRI Imaging, which chooses pointings randomly from a set of 311 possible positions. Details regarding the MIRI imaging dithering patterns can be found in the JWST online documentation. The MIRI double prism, its position called P750L in the filter wheel sequence, disperses 5 -- 14~\um\ light over $\sim$400 pixels, quasi-vertically along the detector columns (Figure \ref{fig:dispSpec}). 

The $\sim$400-pixel length of the spectra means that some sources near the detector edge will only yield partial spectra, with the in-field sources also changing as the telescope dithers during the observation. To work out the origin of partial spectral traces we require a second post-dispersed-sequence direct image; users have the option to obtain two extra imaged dither positions (for a total of 3) up and down in the dispersion direction, to identify sources near but outside the edges of the field whose spectra are visible in the dispersed images (Figure \ref{fig:dispSpec}). 

MIRI does not have dispersing elements that can disperse in orthogonal directions, as is the case for the NIR instruments. If the target of the observation is a crowded field, users can request a second identical sequence to be performed with a position angle offset from the first, to help de-contaminate overlapping spectral traces. 

Observing a field at multiple position angles mitigates spectral overlap in wide-field slitless spectroscopy. Spectra that overlap at one position angle are frequently separated at another, allowing contaminated regions to be identified and the uncontaminated portions of the spectra to be recovered. We developed a tool, {\textit{{{W}ebb's {H}elpful {I}deal-coordinate {P}redictor for {P}ositions, {O}ffsets, and {T}races: WHIPPOT}} \citep{whippot}, to help users visualize the dispersed images of their field of view of interest at different position angles (V3PA in the JWST APT). Figure \ref{fig:whippot} shows an example of using {\textit{WHIPPOT}} for PID 9505 observation planning. 

\begin{figure}
    \centering
    \includegraphics[width=0.9\linewidth]{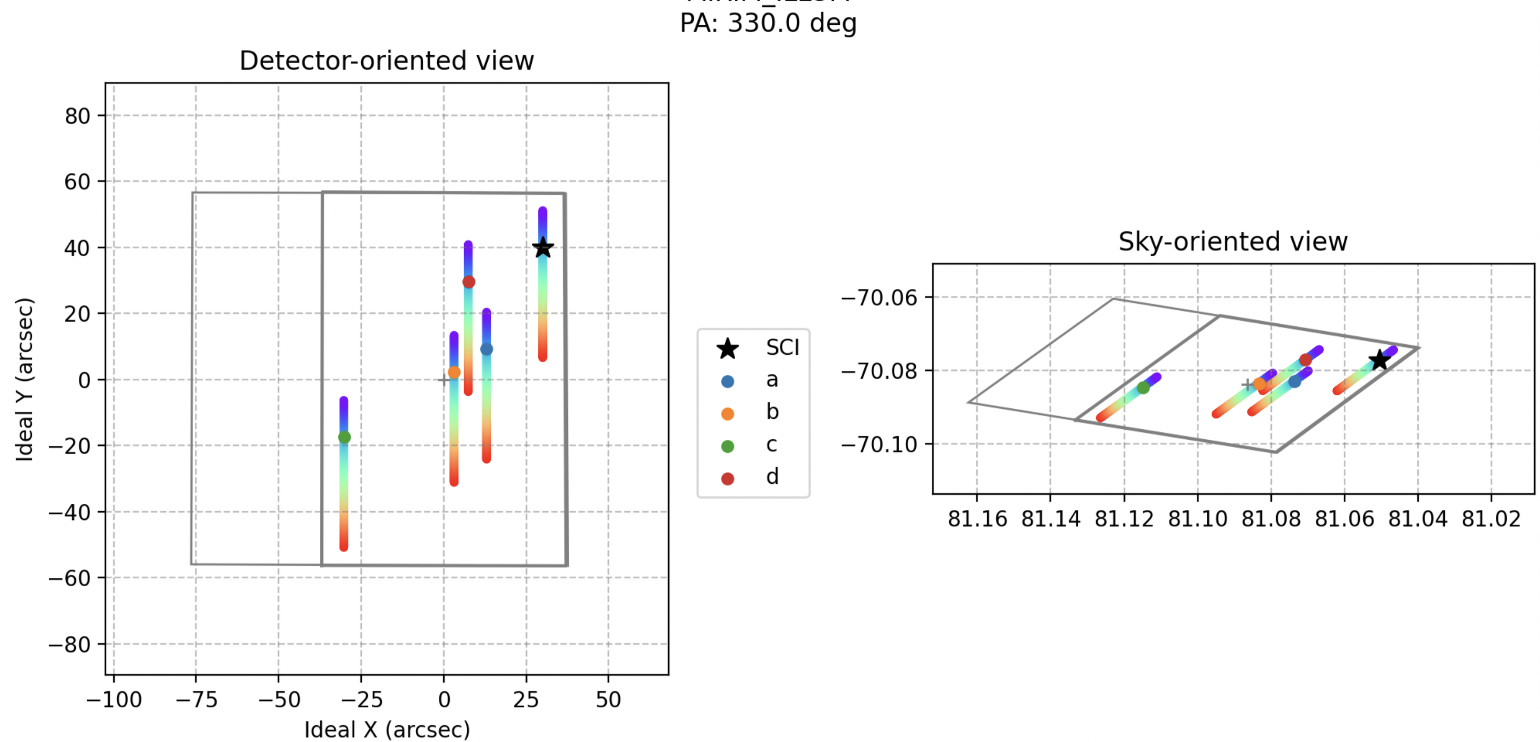}
    \caption{Output schematic from code {\textit{WHIPPOT}} \citep{whippot} used to visualize the spectral traces associated with a particular field of view at a given position angle. For JWST the position angle can be set as a V3PA range in APT.}
    \label{fig:whippot}
\end{figure}

A key challenge in the data calibration for the WFSS mode is the field variation of the dispersion relation of the P750L prism, due to optical distortion. This means the wavelength calibration has to be dynamically adjusted depending on the location of the source. In this scheme, the centroid of a source in the direct image on the detector determines the spectral trace geometry in the dispersed exposures. The JWST calibration pipeline \citep{2019ASPC..523..543B} processes the direct imaging exposures---performing detector calibrations, applying distortion correction and generating the world coordinate system (WCS), and converting to physical flux units---to produce a final catalog  of sources that will be calibrated and extracted from the dispersed exposures following initial calibrations.


If users have a set catalog of source coordinates for which they plan to extract WFSS spectra, then we recommended users perform an independent check of the astrometry of the direct image to ensure that the direct image pixel position corresponds to the source they wish to extract. 

\section{Estimating Count Rates}
The MIRI detectors use up-the-ramp, non-destructive reads, whereby count rates (DN/s) are estimated from linear fits to groups of frames. This process is part of the {\tt calweb\_detector1} pipeline documented in the JWST on-line documentation. The MIRI WFSS pipeline uses the same {\tt calweb\_detector1} as other MIRI modes. 


The JWST {\tt calweb\_detector1} pipeline generates count-rate (\texttt{rate}) products. Before any further analysis, we replace invalid pixels (NaNs) with the median value of neighboring pixels in order to improve the stability of profile fitting and spectral extraction.

\section{Background Subtraction}
The background seen by JWST constitutes several different components~\citep{2023PASP..135d8002R}. In the mid-IR, the background is dominated by the contribution of zodiacal emission to $\sim$12~\um; beyond this wavelength, thermal emission from the observatory begins to dominate~\citep[e.g.,][]{rowlands2025}. The zodiacal background level varies with location on the sky and the date of the observation. The dispersed MIR background as seen by the WFSS mode is the integrated background flux over the full bandpass of the P750L prism. 


The MIRI WFSS pipeline, following the example of other modes, uses a reference background file that is scaled and subtracted from dispersed exposures. We used 813 archival dispersed exposures (LRS slit exposures, in which the imager field is also dispersed) to create the reference background file as follows. From each dispersed image, we subtract a pedestal to account for the changing dark current; i.e., we compute the median for columns 291 to 342 (outside the focal plane mask) and subtract it from each dispersed exposure. We then normalize each dispersed exposure by its median. We then median combine the 813 pedestal-subtracted and normalized dispersed images. The JWST WFSS pipeline subtracts an estimated background emission map from each dispersed image. This is done by normalizing the reference background file to regions free of dispersed signal and then subtracting the scaled background.

Figure \ref{fig:dispSpec} shows a background-subtracted dispersed image. As of this writing, the only optimization performed is the minimization of large residual structures. Improvements in the next few pipeline builds will include corrections for scattered light, as implemented for MIRI MRS observations, and adjustments to the algorithms scaling the reference background file to each exposure by using point source spectral traces to estimate background and source counts simultaneously.

Users can perform their own background subtraction by combining exposures that have sufficiently few artifacts or spectral traces to produce a smooth, location- and date/time-specific background.

\section{Flat}
The MIRI WFSS pipeline currently uses the MIRI imager F770W full-frame flat. Details on the construction of the MIRI imager flats can be found in \citep{shenoy2026}.

We are testing two types of flats for this mode by estimating their impact on the flux calibration by measuring differences between fluxes of a flux calibrator spectral trace at 20 positions on the detector and a CALSPEC model and also by minimizing uncertainties from fits to the signal and background at each detector row (as done when determining the spectral trace; see Section~\ref{sec:FindSTrace}).

The first in-testing flat is built by: subtracting the background from each of the 813 dispersed exposures used to make the background; sigma-clip the result; normalize by the median flux; and median combine the resulting images. 

The second type of in-testing flat is a chromatic (i.e., wavelength dependent) flat where we interpolate the MIRI imager flats to have an estimated responsitivity for each pixel and wavelength. This chromatic flat would then be incorporated in the {\tt extract\_1d} step by estimating the flux at each wavelength in each pixel as the value in counts divided by the flat value at that pixel and wavelength. 

\section{Wavelength Calibration}
 The goal of wavelength calibration is to assign a wavelength to each pixel containing dispersed light from the source of interest. The JWST pipeline achieves this through a reference file known as $\texttt{specwcs}$ that is used to establish the mapping between a source position in the direct image and the detector position of its dispersed light at wavelengths between 5 and 14 $\mu$m.

We aim to minimize systematic errors in precision of wavelength calibration and characterize the accuracy of spectral trace modeling across the field of view. As of this writing this characterization is in progress and is typically not worse than 0.5 pixels, corresponding to 40 nm. We also note that as of this writing we are working on some implementation errors that stem from the number of grid elements required to assign a wavelength vector. Trade-offs between legacy code, speed of reduction, and accuracy are underway. 

The wavelength calibration was derived from observations of sources with known spectral features.  The procedure consisted of detector-level processing, source localization, spectral extraction, line identification, and fitting of a parametric wavelength solution suitable for conversion into a JWST \texttt{specwcs} reference file.

The location of each source was determined from a corresponding direct image. Source positions were then converted to the F770W reference frame by applying filter-dependent offsets using the \texttt{miricoord} software package \citep{miricoord}. Telescope pointing offsets and dither information were subsequently used to determine the source position in the dispersed exposure, denoted $(X_0,Y_0)$. These coordinates represent the expected location of the source in the dispersed image and form the reference point for all subsequent measurements.

\subsection{Spectral Cutout Extraction}

For each point source we use for the MIRI WFSS calibration analysis, a rectangular cutout centered on the dispersed spectrum was extracted from the rate image. The cutout dimensions are $50 \times 420$ pixels. All subsequent analysis is performed in the cutout coordinate system.

The pipeline produces cutouts that have a minimum width of 21 columns. The JWST pipeline spectral cutout sizes are determined by the segmentation maps that connect the direct and dispersed exposures. For partial traces, the portion outside the detector is set to NaN. The construction of spectral cutouts for overlapping MIRI WFSS sources is enabled in the latest pipeline but is currently being tested.




\subsection{Spectral Trace Determination\label{sec:FindSTrace}}

The spectral trace (Figure \ref{fig:spectrace}) was measured by fitting a Gaussian profile independently to each detector row. For every cutout row coordinate $C_Y$, the centroid position $C_X$ was determined from the mean of the fitted Gaussian profile. The resulting sequence of centroids defines the trace. This trace model provides the expected horizontal position of the spectrum as a function of detector row.


\begin{figure}
    \centering
    \includegraphics[width=0.9\linewidth]{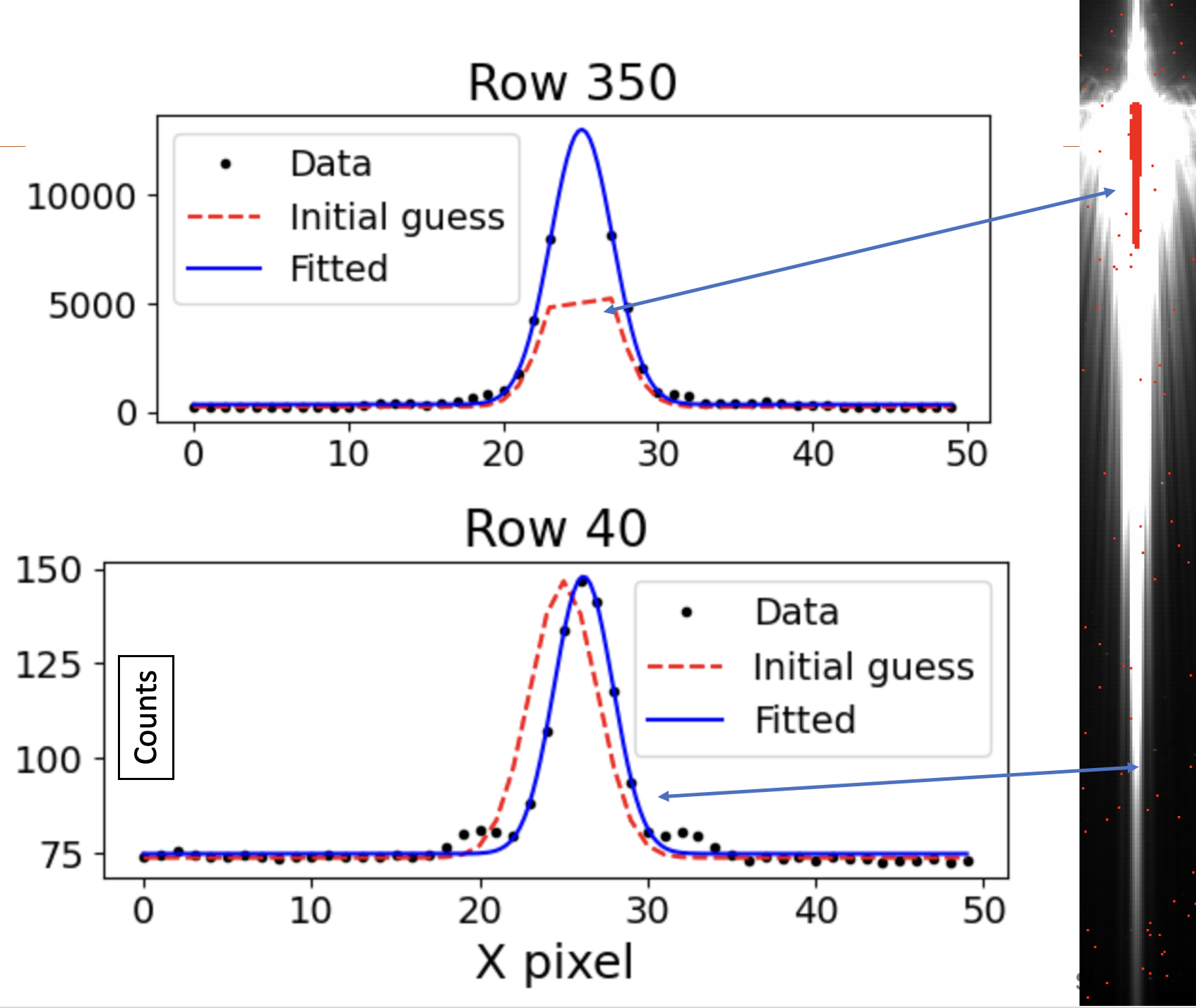}
    \caption{This figure illustrates our method of finding the centroid of dispersed emission on the detector by fitting a Gaussian and a background component at each row.}
    \label{fig:spectrace}
\end{figure}

\subsection{One-Dimensional Spectrum Extraction}

A one-dimensional spectrum is then extracted from the cutout image from integration of the fitted Gaussian profiles at each row along the dispersion direction. The trace model (i.e., x- and y-offsets from the direct image position as shown in Figure \ref{fig:DXDYLam}) was then used to determine the corresponding offsets for the centroid of emission at the center of each identified spectral line $C_X$, yielding a set of calibration measurements.

\begin{figure}
    \centering
    \includegraphics[width=0.99\linewidth]{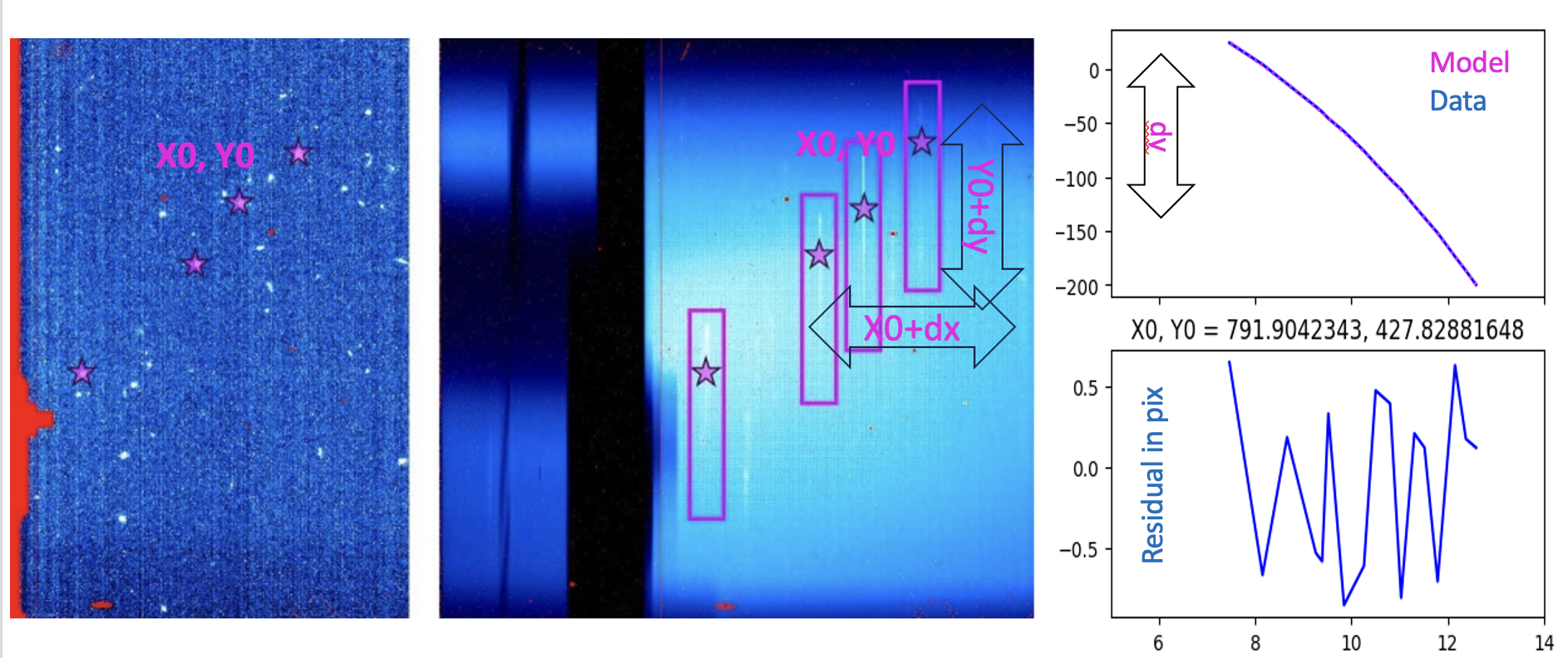}
    \caption{The left-most panel shows the direct image associated with a WFSS observation; the center panel shows the dispersed image with the x- and y-offsets marked with pink arrows; the figures at the right show the difference between the modeled dispersion in pink and the data in blue.}
    \label{fig:DXDYLam}
\end{figure}

\subsection{Spectral Line Identification}

\begin{table*}[t]
\centering
\caption{\large Wavelength Calibration observations}
\label{tab:cal_lines}
\renewcommand{\arraystretch}{1.15}
\begin{tabular}{lll}
\hline
Source & Observations & Emission lines ($\mu$m) \\
\hline

LHA-120-N-133 &
\begin{tabular}[c]{@{}l@{}}
PID 9505\\
4 exposures\\
6 groups/int\\
10 integrations
\end{tabular}
&
6.27, 7.89, 8.66, 9.00, 10.52, 11.28, 12.82 \\

\hline

HD76534 &
\begin{tabular}[c]{@{}l@{}}
PID 9266\\
20 exposures\\
5 groups/int\\
36 integrations
\end{tabular}
&
\begin{tabular}[c]{@{}l@{}}
7.47, 8.16, 8.66, 9.26, 9.39, 9.52,\\
9.85, 10.26, 10.50, 10.80, 11.03, 11.31,\\
11.52, 11.79, 12.15, 12.38, 12.59
\end{tabular}
\\

\\

\hline

Archival sources &
\begin{tabular}[c]{@{}l@{}}
PID 4533\\
2 sources\\
3 detector positions\\
2 exposures\\
97 groups/int\\
4 integrations
\end{tabular}
&
Wavelength coverage: 6.5--11.4 \\

\hline
\end{tabular}
\end{table*}

Candidate emission features were identified using the \texttt{SciPy find\_peaks} peak-finding algorithm applied to the extracted one-dimensional spectrum. Detected peaks were compared with a reference spectrum possessing known wavelength assignments. Ambiguous or blended features were removed through visual inspection, producing a final list of calibration lines (Figure \ref{fig:RefinedPeak}). For each accepted line, the detector row coordinate $C_Y$ was associated with a wavelength $\lambda$. To refine $C_Y$ to subpixel values we did a subsequent Gaussian fit for each spectral line with $C_Y$ as the initial guess for the peak and we included a background line fit. Each fit was checked visually and we recorded each $\chi^2$/degree of freedom (DOF) to prevent overfitting (i.e., having too few data points for the parameters we are trying to fit would give a $\chi^2$/DOF $\leq$ 1).

\begin{figure}
    \centering
    \includegraphics[width=0.54\linewidth]{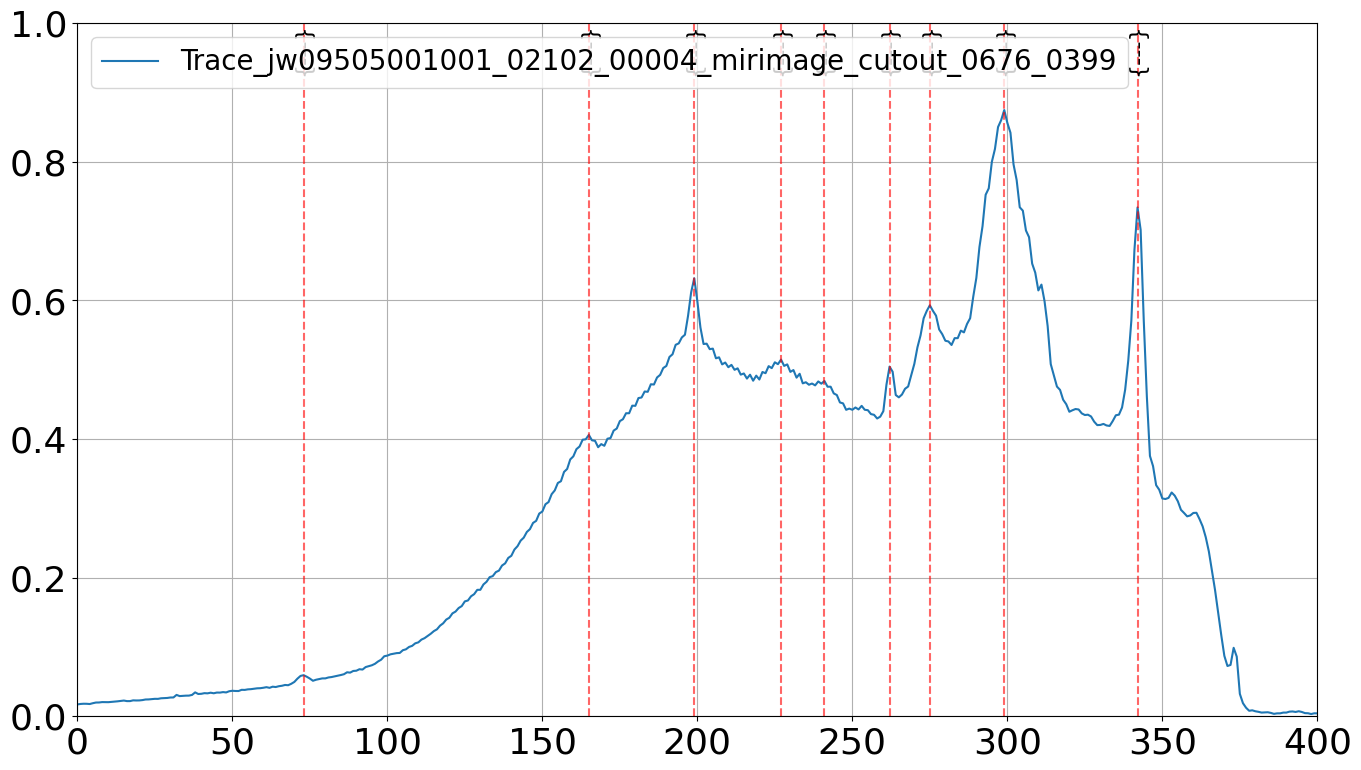}
    \includegraphics[width=0.45\linewidth]{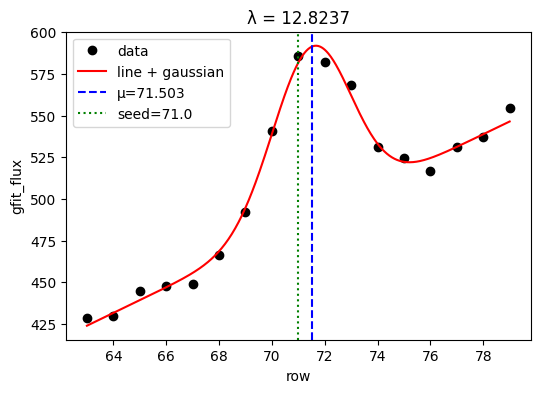}\\
    \caption{(left) Example of spectral features used for wavelength calibration. Emission lines are identified automatically (left panel) and then refined using a Gaussian profile and a linear continuum fit (right panel).}
    \label{fig:RefinedPeak}
\end{figure}

The calibration measurements were converted into the coordinate system required by the JWST spectral configuration framework. For each calibration line, detector offsets $dy$ in the dispersion direction and $dx$ in the cross-dispersion direction relative to the source position were computed. The complete calibration dataset therefore consists of measurements of ($x_0$, $y_0$, $dx$, $dy$, $\lambda$). These quantities describe the location of monochromatic light on the detector as a function of source position and wavelength. We then combined the 24 spectral traces to obtain a predictive model for the spectral trace at each position. 






\subsection{Polynomial Wavelength Model}

The wavelength dependence of the trace was modeled using the polynomial formalism adopted by the JWST spectral configuration files \citep[e.g.,][]{norruss17}. A normalized wavelength parameter,

\begin{equation}
t =
\frac{\lambda - \lambda_{\min}}
{\lambda_{\max} - \lambda_{\min}},
\end{equation}

was introduced such that $t=0$ corresponds to the shortest wavelength in the calibration range and $t=1$ corresponds to the longest wavelength.

The detector offsets were represented as generalized two-dimensional polynomials in source position whose coefficients vary with wavelength and the position of the centroid in the direct image. We continue to explore the degree and type of polynomials that best characterize the spatial dependency of the spectral trace and allow accurate and computationally efficient wavelength calibration. The polynomial coefficients were determined through least-squares fitting to the calibration measurements.





The fitted polynomial coefficients were written to a spectral configuration file describing the trace geometry and wavelength mapping. The resulting model was tested by comparing predicted and measured line locations across the calibration dataset. Following validation, the configuration file was converted into a JWST \texttt{specwcs} ASDF file in the Calibration Reference File Data Systems (CRDS). Reference files are updated regularly and, for this new mode in particular, users should check if their pipeline version includes the most recent reference files. 

\section{Flux Calibration}

Flux calibration converts counts to physical units which for MIRI are given as (MJy/sr). The JWST pipeline uses a \texttt{photom} reference files that gives a vector of the relative responsivity as a function of wavelength and a scaling in units of $\frac{\mathrm{MJy}\,\mu\mathrm{m}}{\mathrm{DN}\,\mathrm{s}^{-1}}$. Our approach leverages the LRS slit and slitless subarray modes, whose flux calibration is described in detail in \citep{wong2025}. However, because the spectral traces change curvature across the imaging aperture as described in Section \ref{sec:FindSTrace}, all WFSS modes use a wavelength-dependent {\tt photom} file. 

Because the point-spread function across the detector as a function of wavelength is not yet well-characterized, we do not use an aperture correction; instead, we start with the LRS photom file, scaling the MIRI WFSS data match the flux and continuum shape of the MRS and LRS data of the science verification target LHA-120-N-133.

Figure \ref{fig:FcalCenter} compares the resulting WFSS and LRS spectra of LHA-120-N-133. While the differences between the LRS and WFSS spectra are on the order of the noise in both ($\leq 1\%$), these results were obtained for a source at the center of the imager and for the source used to optimize the calibration. Characterization and improvement of the flux calibration across the field of view are in progress, with additional flux and wavelength calibration observations planned for later in 2026.

\begin{figure}
    \centering
    \includegraphics[width=0.8\linewidth]{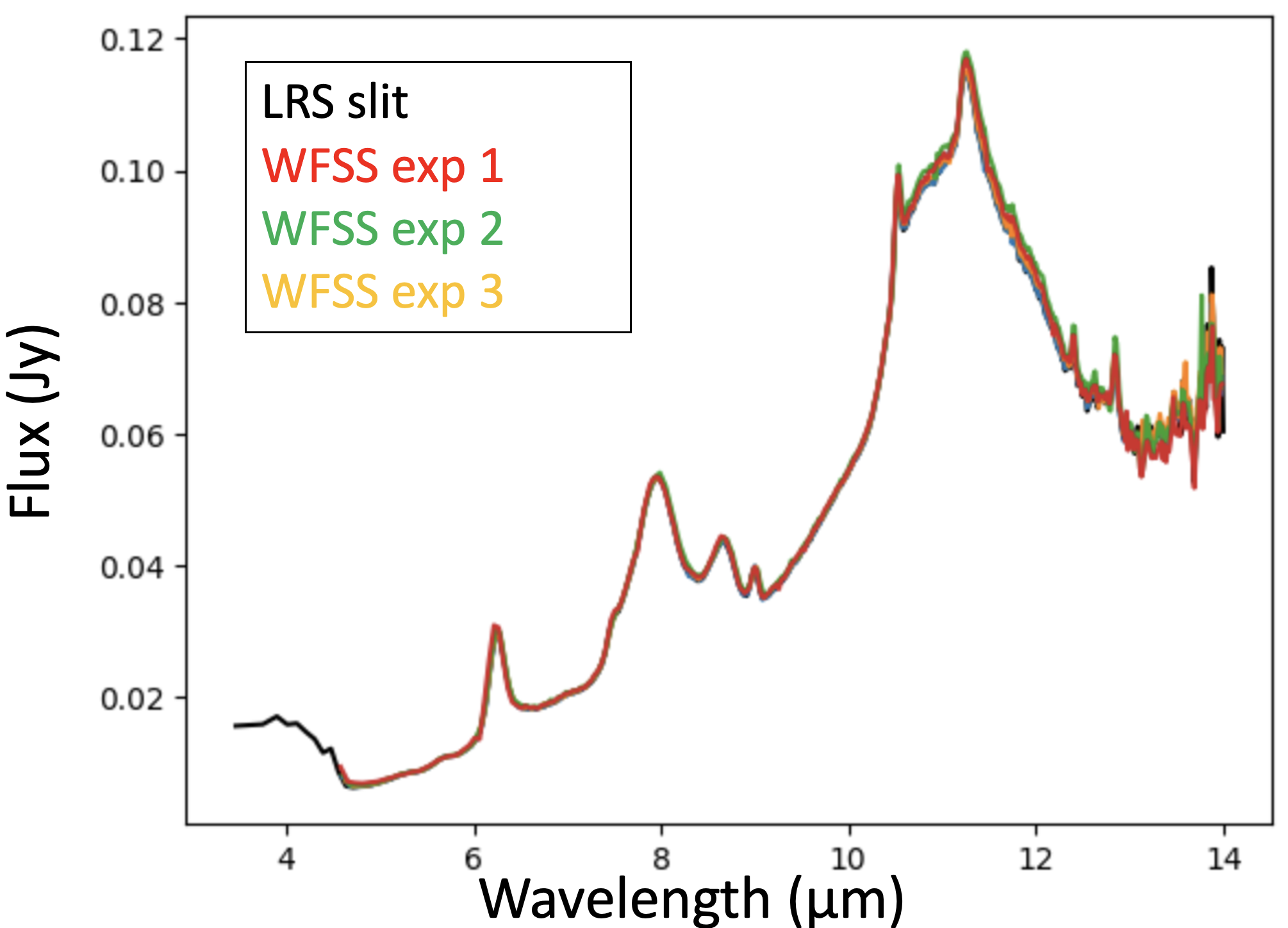}
    \caption{Comparison between LRS slit (PID 4491) and WFSS spectra (PID 9505) of LHA-120-N-133, a planetary nebula observed as part of WFSS science verification.}
    \label{fig:FcalCenter}
\end{figure}

\section{Deblending Overlapping Spectral Traces}



The JWST pipeline also includes algorithms to deblend overlapping spectral traces, documented in the JWST pipeline Read the Docs pages for {\textit{wfss\_contam}}, and briefly illustrated in Figure \ref{fig:NMcontam}. Specifically, at each pixel that the segmentation map considers to be part of the source of interest, the direct image is dispersed into the grism frame using the {\textit{specwcs}} model and inverse flux-calibrated back into DN/s. A contamination model is generated by dispersing all other sources from the direct image in a similar manner and differences between model and observations are minimized. We are currently testing and optimizing this algorithm for MIRI WFSS.

\begin{figure}
    \centering
    \includegraphics[width=0.9\linewidth]{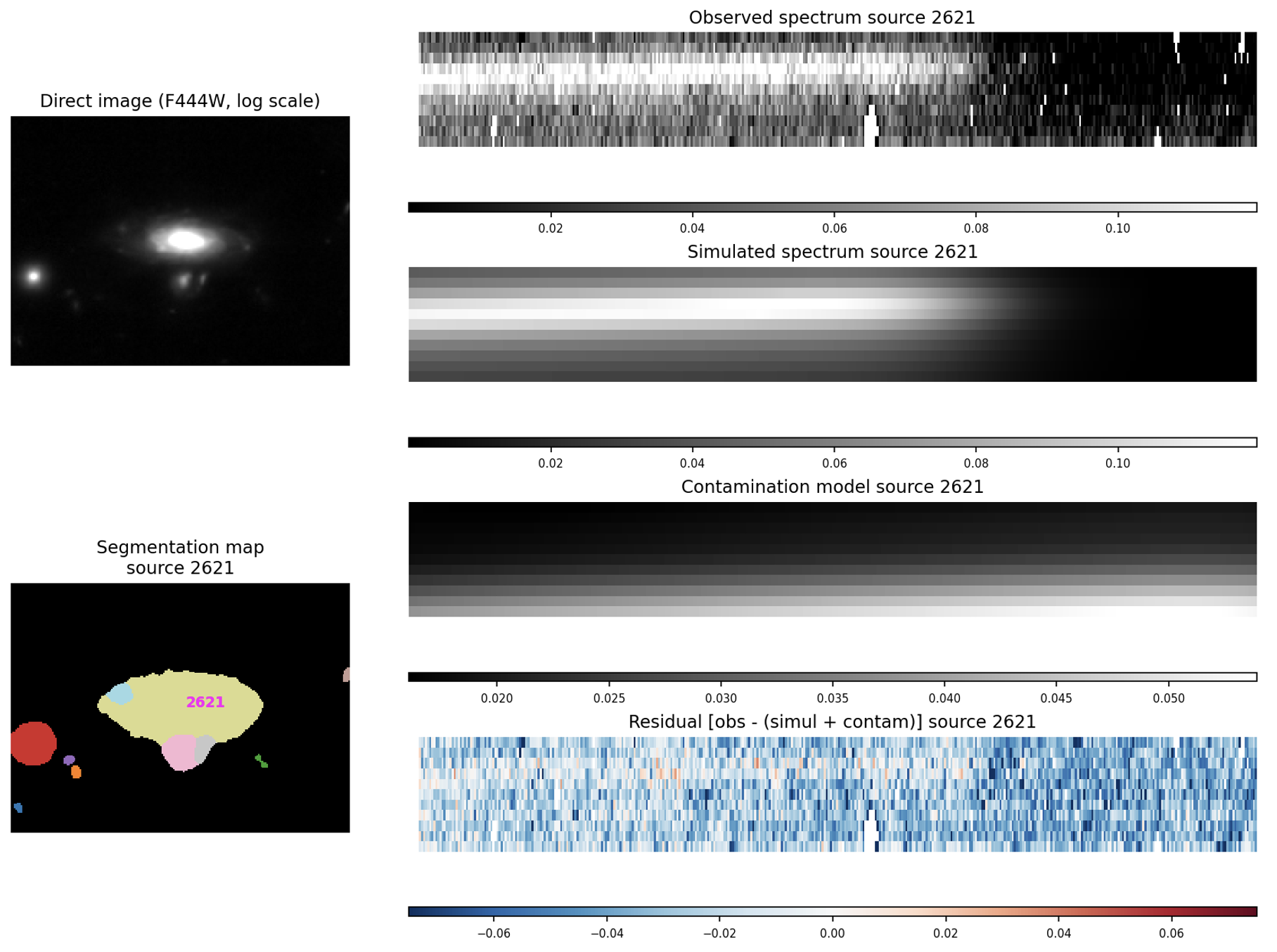}
    \caption{The top left panel shows the fully-calibrated direct image, cut out around the source of interest. The bottom left panel shows the same cutout region for the segmentation map, which is generated by running a watershed segmentation algorithm on the direct image (with deblending) using the {\tt SourceCatalogStep} (backed by {\tt photutils}). 
    The simulated spectrum is shown in the ``Simulated spectrum'' panel on the right. We can see it compares well with the ``Observed spectrum'' panel. 
    Finally, the residual is generated by adding the simulated spectrum and contamination model, and subtracting their combination from the observed spectral trace. } 
    \label{fig:NMcontam}    
\end{figure}

\section{Conclusions}
The MIRI WFSS pipeline now produces segmentation maps, background-subtracted dispersed images, and wavelength- and flux-calibrated extracted 2D and 1D spectra. The wavelength calibration achieves the required accuracy for science observations at the center of the field of view, while the flux calibration is based on the LRS slit calibration, modified to account for field-dependent changes in dispersion and refined by comparison with MRS observations of a calibration source.

Calibration and pipeline development are ongoing. Planned improvements include additional dedicated calibration observations, the incorporation of archival spectra of point sources with known redshifts, and evaluation of alternative models for the spectral traces to further improve the wavelength and flux calibrations.

Cycle 6 will include the opportunity for coordinated and pure parallel observations with NIRCam which will be highly synergistic to planned surveys with {\textit{Roman}} and {\textit{SPHEREx}}.
 

\bibliography{report} 
\bibliographystyle{spiebib} 

\end{document}